\documentclass{aa}
\usepackage{latexsym}
\usepackage{natbib}
\usepackage{graphicx}
\usepackage{times}  \DeclareSymbolFont{operators}{OT1}{ptm}{m}{n}
\usepackage{color}
\newcounter{IonCS}
\renewcommand{\ion}[2]{\setcounter{IonCS}{#2}#1\,{\scshape{\roman{IonCS}}}}

\newcommand{\sect}[1]{Sect.\,\ref{#1}}

\newcommand{\fig}[1]{Fig.\,\ref{#1}}
\newcommand{\figs}[1]{Figs.\,\ref{#1}}

\newcommand{\eqn}[1]{Eq.\,(\ref{#1})}
\newcommand{\app}[1]{Appendix\,\ref{#1}}
\graphicspath{{./}{figs-ps/}{figs/}}

\begin{document}

%
\title{Structure of solar coronal loops: from miniature to large-scale}

\titlerunning{Structure of solar coronal loops}
\authorrunning{H. Peter et al.}

\author{H.~Peter\inst{1}, S. Bingert\inst{1}, 
        J. A. Klimchuk\inst{2},
        C. de\,Forest\inst{3}, 
        J. W. Cirtain\inst{4}, 
        L. Golub\inst{5}, 
        A. R. Winebarger\inst{4}, 
        K. Kobayashi\inst{6},
        K. E. Korreck\inst{5}
        }

\institute{Max Planck Institute for Solar System Research,
           37191 Katlenburg-Lindau, Germany, email: peter@mps.mpg.de
           \and
           Heliophysics Division, NASA Goddard Space Flight Center, Greenbelt, MD 20771, USA
           \and
           Southwest Research Institute, Instrumentation and Space Research Division,
           Boulder, CO 80302, USA.
           \and
           Marshall Space Flight Center, NASA, Mail Code ZP13, MSFC, Alabama 35812, USA
           \and
           Harvard-Smithsonian Center for Astrophysics, 60 Garden Street, Cambridge, Massachusetts 01238, USA
           \and
           Center for Space and Aeronautic Research, University of Alabama, Huntsville, Alabama 35812, USA
}

\date{Received 3 May 2013 / Accepted 19 June 2013}

\abstract%
%
{%
}
{%
We will use new data from the High-resolution Coronal Imager (Hi-C) with unprecedented spatial resolution of the solar corona to investigate the structure of coronal loops down to 0.2\arcsec.
}
{%
During a rocket flight Hi-C provided images of the solar corona in a
wavelength band around 193\,\AA\ that is dominated by emission from
\ion{Fe}{12} showing plasma at temperatures around 1.5\,MK. We
analyze part of the Hi-C field-of-view to study the smallest coronal
loops observed so far and search for the a possible sub-structuring
of larger loops.
 }
{%
We find tiny 1.5\,MK loop-like structures that we interpret as miniature
coronal loops. These have length of the coronal segment above the
chromosphere of only about 1\,Mm and a thickness of less than 200
km. They could be interpreted as the coronal signature of small flux
tubes breaking through the photosphere with a footpoint distance
corresponding to the diameter of a cell of granulation. We find loops that are longer than 50\,Mm to have a diameter of about 2\arcsec\ or 1.5\,Mm,
consistent with previous observations. However, Hi-C really resolves
these loops with some 20 pixels across the loop. Even at this
greatly improved spatial resolution the large loops seem to have no
visible sub-structure. Instead they show a smooth variation in
cross-section. }
{%
The fact that the large coronal loops do not show a sub-structure at the spatial scale of 0.1\arcsec\ per pixel implies that either the densities and temperatures
are smoothly varying across these loops or poses an upper limit on the diameter of strands the loops might be composed of. We estimate that strands that compose the 2\arcsec\ thick loop would have to be thinner than 15\,km.
The miniature loops we find for the first time pose a challenge to be properly understood in terms of modeling.
}
%

\maketitle

\section{Introduction\label{S:intro}}

The basic building blocks of the corona of the Sun are coronal loops covering a wide range of temperatures. Their lengths cover a vast range from only a few Mm to a sizable fraction of the solar radius. Loops have been revealed as early as the 1940s in coronagraphic observations \citep[][Sect 1.4]{Bray+al:1991} and then in X-rays \citep[e.g.][]{Poletto+al:1975} showing the close connection of the hot coronal plasma of several $10^6$\,K to the magnetic field. When investigating cooler plasma at around $10^6$\,K, e.g. in the spectral bands around 171\,\AA\ and 193\,\AA\ dominated by emission from \ion{Fe}{9} and \ion{Fe}{12}
formed at 0.8\,MK and 1.5\,MK, the loops show up at high contrast.
Small loops related to the chromospheric network are seen at transition region temperatures of about 0.1\,MK, e.g. in \ion{C}{3} or \ion{O}{6} \citep[e.g.][]{Peter:2001:sec,Feldman+al:2003}. In all these observations, the sub-resolution spatial and thermal structure of the loops remains unknown.

The highest resolution data from the corona we currently get on a regular basis are from the Atmospheric Imaging Assembly \cite[AIA;][]{Lemen+al:2012}
on the Solar Dynamics Observatory \citep[SDO;][]{Pesnell+al:2012}. With its spatial scale of 0.6\arcsec\ per pixel and a spatial resolution slightly worse than 1\arcsec\ at least part of the 1\,MK loops show a smooth cross section and seem to be resolved \citep{Aschwanden+Boerner:2011}. These results hint at the loops having a narrow distribution of temperatures, like other previous spectroscopic studies revealed \citep{DelZanna+Mason:2003}, even though they might not be truly isothermal. Therefore some spatial substructure should be expected \citep[e.g.][]{Warren+al:2008}. Recently in their analysis of AIA data \cite{Brooks+al:2012} placed a limit on the diameter of strands composing a loop of some 200\,km and more. Likewise \cite{Antolin+Rouppe:2012} argued that coronal loops should have substructures of 300\,km or smaller, which is based on observations of coronal rain. A detailed discussion of observations and modeling of multi-stranded loops can be found in the review of \cite{Reale:2010}.

To model the (sub) structure of loops one can assume that one single loop is composed by a number of individual strands. In the first  of such models \cite{Cargill:1994} used 500 strands, each of which was heated impulsively by nanoflares following the concept of \cite{Parker:1983,Parker:1988}. Many of such models have been investigated since, modifying various parameters with the final goal to empirically understand the appearance of large loops \citep[for a recent approach see e.g.][]{LopezFuentes+Klimchuk:2010}. In 3D MHD models one can directly study the structure of coronal loops, in particular the relation of the coronal emission to the magnetic field. These show that the cross section of the magnetic structure hosting the loop is non-circular and changing along the loop \citep{Gudiksen+Nordlund:2005b,Mok+al:2008}
and that the resulting loop seen in coronal emission appears to have a constant cross section \citep{Peter+Bingert:2012}. Both \cite{Mok+al:2008} and \cite{Peter+Bingert:2012} emphasized that it is not \emph{only} the magnetic structure that defines the loop visible in coronal emission: one has to consider carefully also the thermal structure that forms along and across the magnetic structure. The spatial resolution of these 3D models is (as of yet) not sufficient to study the substructure of loops, i.e. to see if strands form within a loop, where the diameter of these strands is smaller than current observations allow to resolve.

Because the nature of the internal structure of loops is of high interest to the heating mechanism, it is of importance to investigate if the loops are monolithic or multi-stranded --- and if they are multi-stranded to place limits on the diameter of the strands. Likewise, it is of importance to identify and investigate the smallest structures radiating at coronal temperatures. Is there a lower limit for the length of a coronal loop, or are there short structures hidden below the resolution limit of current instrumentation? To address these questions we investigate observations from the High-resolution Coronal Imager \citep[Hi-C;][]{Cirtain+al:2013}. Even though being flown on a sub-orbital rocket and providing only a few minutes worth of data, the spatial resolution is almost six times better than with AIA. This allows us to place new (upper) limits on the strand diameter and to identify miniature coronal loops which are significantly smaller than observed before (at least by a factor of 10) --- smaller than even the tiny cool loops related to the chromospheric network in the transition region.

\begin{figure}
\includegraphics{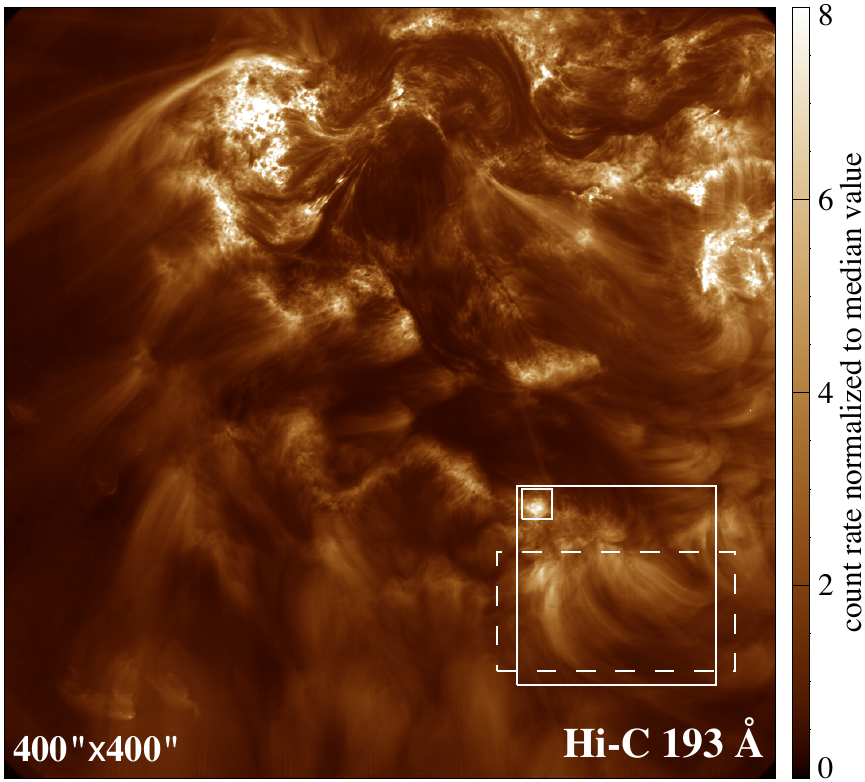}
\caption{Full field-of-view of the Hi-C observations. This image is
taken in a wavelength band around 193\,{\AA} that under active
region conditions is dominated by emission from \ion{Fe}{12} formed
at around 1.5\,MK. The core of the active region with several
sunspots is located in the top half of the image (cf.\
\fig{F:chromo}). Here we concentrate on the loop
system in the bottom right at the periphery of the active region.
The regions indicated by the large solid square and the dashed
rectangle are shown in \figs{F:roi} and \ref{F:model}. The small
square shows the plage areas zoomed-in in \fig{F:plage}. Here as in
the following figures the count rate is plotted normalized to the
median value in the respective field-of-view. North is top.
\label{F:full}}
\end{figure}

The paper is organized as follows. In \sect{S:strategy} we give an introduction to the observations with Hi-C and their relation to the data from SDO in the Hi-C field of view. The miniature loops are discussed in \sect{S:plage} before we investigate the substructure of larger loops (\sect{S:loops}) and the upper limit of the strands (\sect{S:strands}), should loops not be monolithic. We then briefly turn to a comparison of the structures seen in Hi-C to those found in a 3D MHD model in \sect{S:model}, before we conclude our study.

\section{Observations: Hi-C and AIA\label{S:strategy}}

Images of the corona with unprecedented spatial resolution have been obtained during a rocket flight by the High-resolution Coronal Imager (Hi-C). The instrument and first results have been described by \cite{Kobayashi+al:2013} and \cite{Cirtain+al:2013}. The Hi-C experiment recorded data in a wavelength band around 193\,\AA\ with 2\,s exposure time. Under active region conditions this is dominated by emission from \ion{Fe}{12} (193\,\AA) originating at about 1.5\,MK. The spatial scale is about 0.1\arcsec\ per pixel corresponding to 73 km/pixel. This is about a factor of 5.8 better than what is achieved by AIA/SDO, which is the current workhorse for solar coronal extreme UV imaging studies. The temperature responses of the 193\,\AA\ channels of Hi-C and AIA are very similar.
The effective area of HiC is approximately 5.3 times larger than that of AIA, though the Hi-C pixels cover a roughly 36 times smaller area on the Sun.

\begin{figure}
\includegraphics{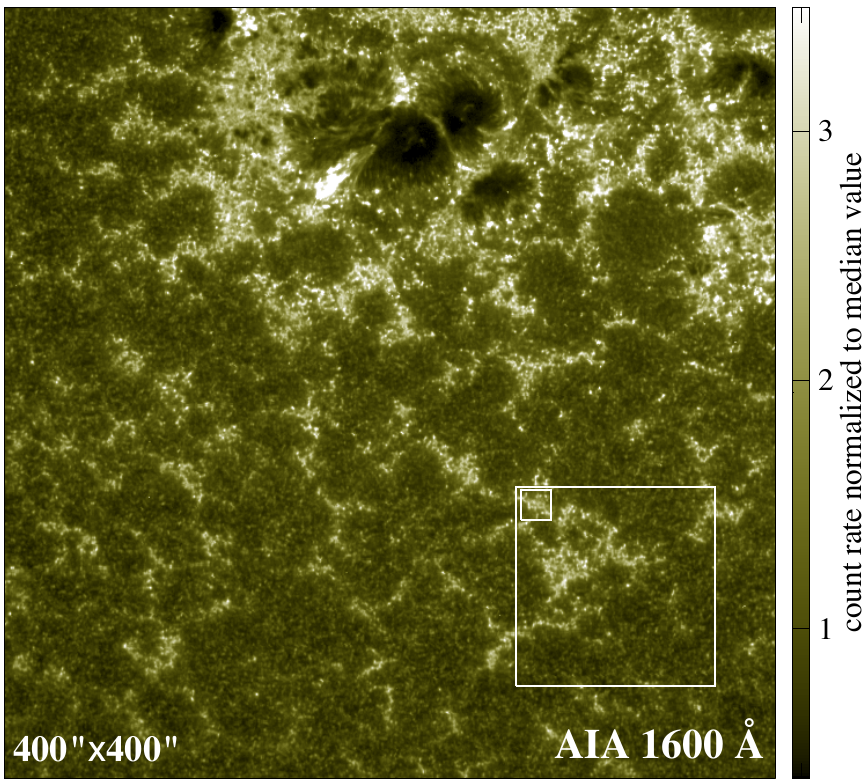}
\caption{Image of the chromosphere co-spatial and co-temporal with
\fig{F:full} taken by AIA in the 1600\,\AA\ channel. As in
\fig{F:full} the large and small squares indicate the field-of-view
displayed in \fig{F:roi} and the zoom of the plage region in
\fig{F:plage}.
\label{F:chromo}}
\end{figure}

In \fig{F:full} we show the full field-of-view of the Hi-C
observation. This frame, which we will investigate in this study,
has been taken at around 18:54:16 UT on 11 July 2012. In this study
we will concentrate on the clear loop-structures in the lower right
part of that image. This is in the periphery of the active region,
away from the sunspots that are found in the upper half of the image
(cf.\ \fig{F:chromo}).

\begin{figure*}
\includegraphics{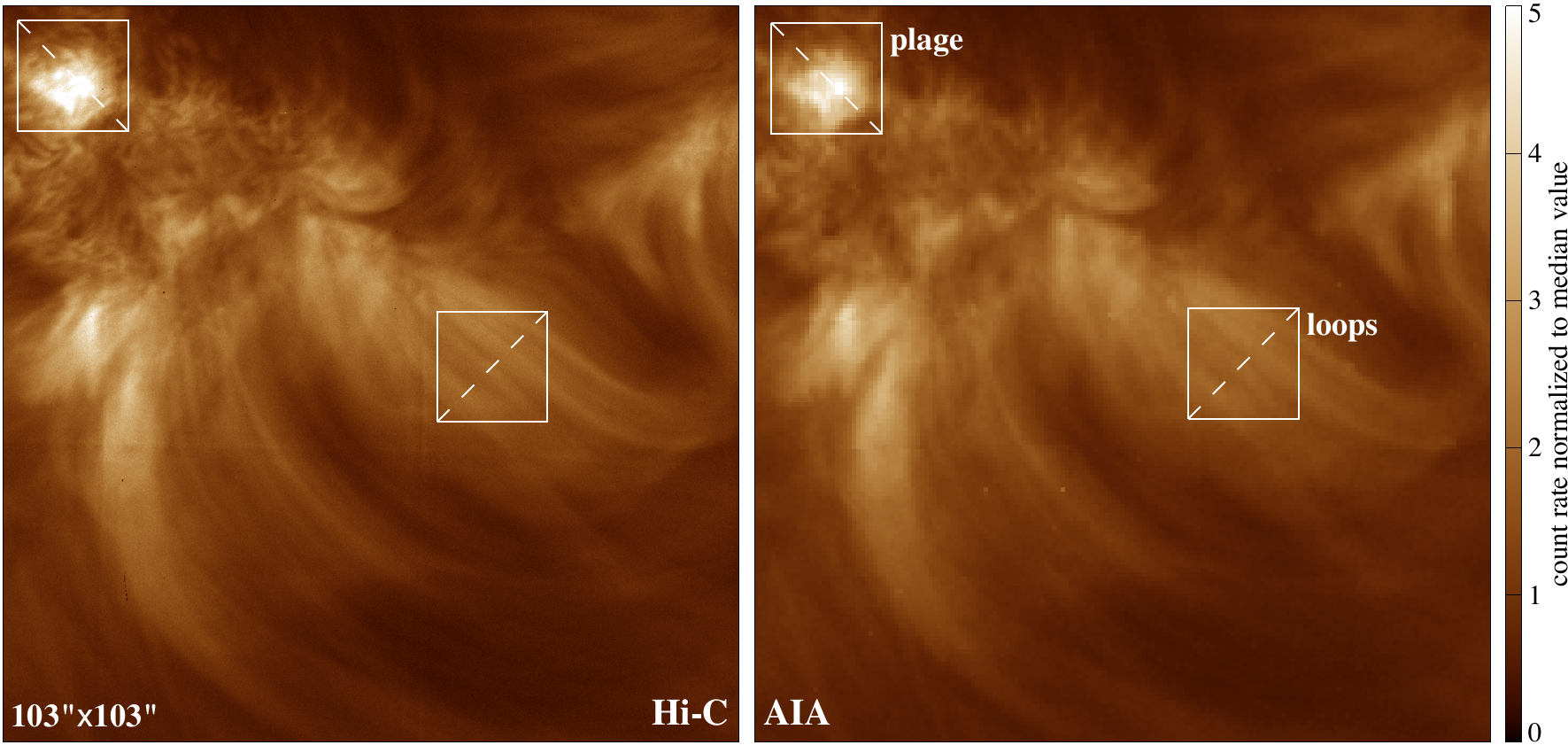}
\caption{Loop system at the periphery of the active region. This shows the 103\arcsec$\times$103\arcsec\ region indicated by the square in \fig{F:full}. The left panel shows the Hi-C observation (1000$\times$1000 pixels), the right displays the data in the same wavelength channel (193\,\AA) recoded by AIA (173$\times$173 pixels). The AIA image is spatially aligned with the Hi-C image and was taken at roughly the same time. The two squares here indicate regions that are magnified in \figs{F:plage} and \ref{F:loop} and that are located in areas dominated by a plage region and a loop system.
\label{F:roi}}
\end{figure*}

The goal of this study is to investigate coronal features that are
not resolvable with AIA. We thus compare the Hi-C image to an AIA
image taken in the same 193\,\AA\ band only seconds after the Hi-C
image. To align the images we had to compensate for a rotation of
1.9$^\circ$ and found the (linear) pixel scaling from Hi-C to AIA to
be a factor of 5.81. For our analysis we will also use images from
the other AIA channels, all of which have been taken between 3\,s
before and 6\,s after the Hi-C image. These we spatially scaled and
aligned to match the AIA\,193\,\AA\ image, and then applied the same
rotation as for the 193\,\AA\ image to have a set of co-spatial
images from Hi-C and AIA. We also make use of the magnetogram taken
by the Helioseismic and Magnetic Imager
\citep[HMI/SDO;][]{Scherrer+al:2012} at 18:53:56, just 20\,s before the
Hi-C image. We scale, rotate and align the magnetogram to match the
AIA 1600\,\AA\ image, so that it is also co-spatial with the Hi-C
image.

\section{Coronal structures\label{S:structures}}

A rough inspection of the region-of-interest for our study as shown in \fig{F:roi} already shows that parts of the Hi-C image look much more crisp than the AIA 193\,\AA\ image, a clear effect of the improved spatial resolution of Hi-C. However, other parts of the image look quite alike, which is particularly true for the large loops in the middle of the region-of-interest.

\subsection{Miniature coronal loops in plage region\label{S:plage}}

\begin{figure}
\includegraphics{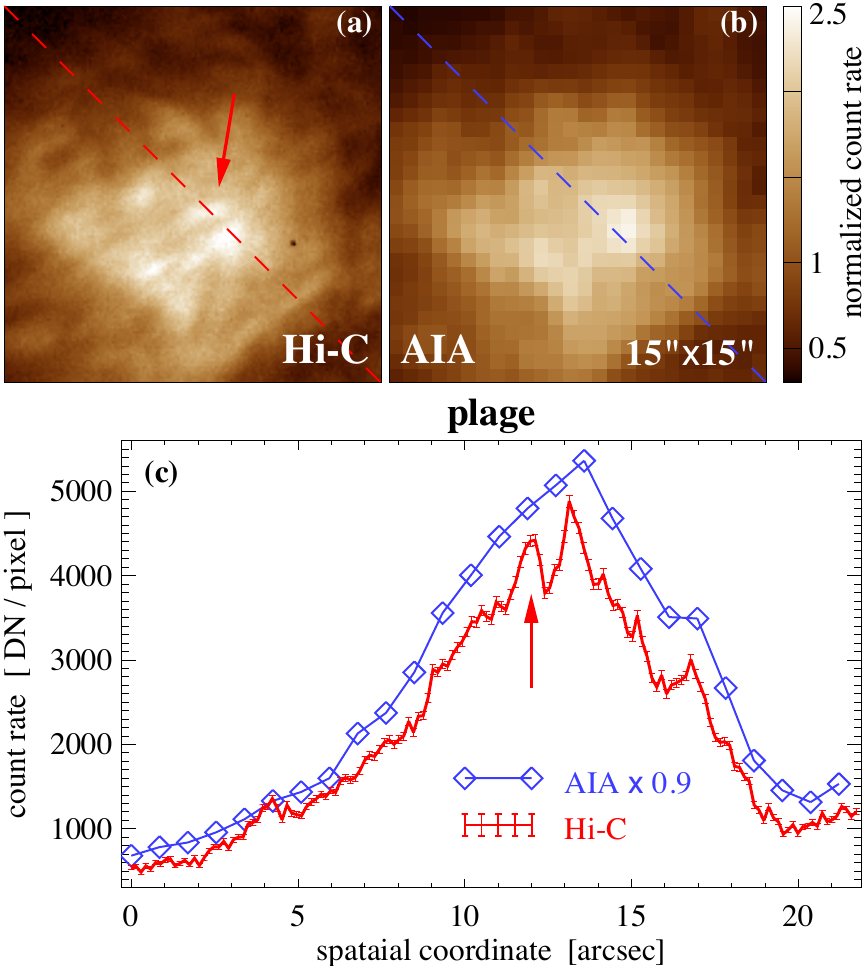}
\caption{Zoom of the \emph{plage region} indicated in \fig{F:roi} by a square. The top two panels show the Hi-C image (150$\times$150 pixel) and the AIA image (26$\times$26 pixel) in that 15\arcsec$\times$15\arcsec\ region. The individual pixels are clearly identifiable in the AIA image. The bottom panel shows the variation of the count rate across the structures along the diagonal indicated by the dashed lines in the top panels. The pixels for the AIA data are indicated by diamonds. The bars indicate the individual pixels of Hi-C, the height of the bars represent the errors. For better comparison the AIA count rate is scaled by the factor given in the plot. The arrows in panels (a) and (c) indicate the position of a miniature coronal loop.
See \sect{S:plage}
\label{F:plage}}
\end{figure}

To highlight that Hi-C shows miniature coronal loop-like structures almost down to its resolution limit we first investigate a small region in the upper left of the region-of-interest, labeled ``plage'' in \fig{F:roi}. A zoom of this area is shown in the top panels of \fig{F:plage}. Here the increased spatial resolution of Hi-C is clearly evident. To emphasize this, panel (c) of \fig{F:plage} shows diagonal cuts through the Hi-C and AIA images in panels (a) and (b). Prominent substructures are seen in the Hi-C image that are some seven to ten Hi-C pixels wide, corresponding to below 1\arcsec\ (e.g.\ the one indicated by the arrow) -- this corresponds to 1.5 AIA pixels and is therefore not resolved by AIA. Besides these, also various smaller intensity peaks are visible, which are only 2 pixels wide, with the intensity enhancement being clearly above the error level\footnote{%
To estimate the error we assumed Poisson statistics of the photon counting, and propagated this error to the count rate (0.244 photons per DN) accounting for a read-out noise of 18.5 DN. 
}.
This confirms that Hi-C can detect small structures down to its resolution limit.

\begin{figure}
\includegraphics{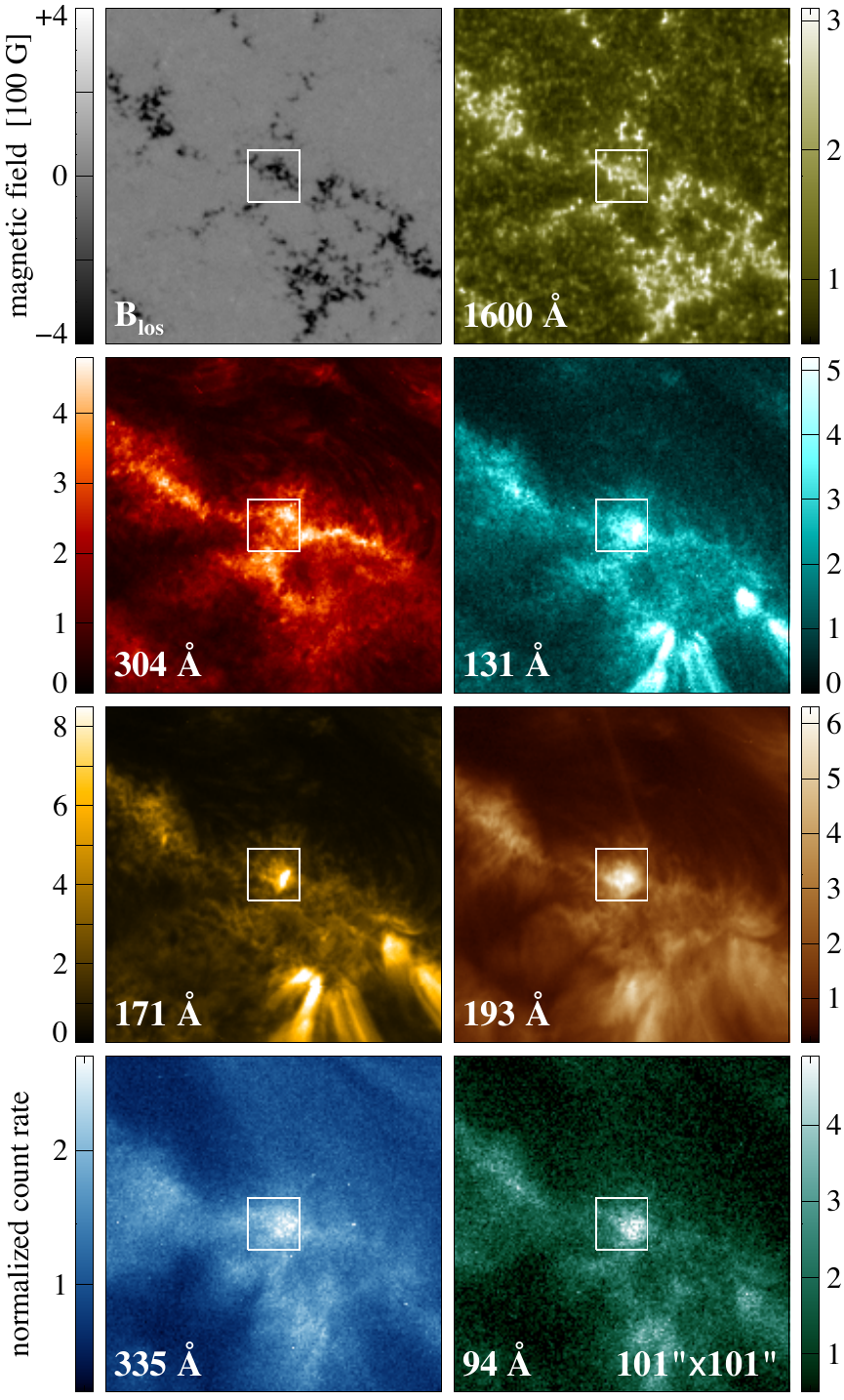}
\caption{Context of the plage region shown in \fig{F:plage}. The AIA images in 7 channels show a 101\arcsec${\times}$101\arcsec\ region and are taken within seconds of the 193\,\AA\ Hi-C image. The squares indicate the field-of-view show in \fig{F:plage} and are co-spatial with the small squares in \fig{F:roi} labeled ``plage''. In \fig{F:plage} the 193\,\AA\ images of Hi-C and AIA zooming into the small square are shown. Only there the miniature loop can be identifies in the Hi-C image.
The top left panel shows the co-spatial line-of-sight magnetogram in the photosphere as seen by HMI, taken within seconds of the other images.
See \sect{S:plage}.
\label{F:context.plage}}
\end{figure}

The nature of the small-scale intensity enhancements in the Hi-C
193\,\AA\ band needs some investigation. In \fig{F:context.plage} we
plot the context of this small plage region in various AIA channels
to investigate the connection throughout the atmosphere. The region
of the strong brightening in 193\,\AA\ coincides with enhanced
emission from the chromosphere as seen in the AIA 1600\,\AA\ channel
(dominated by the \ion{Si}{1} continuum). Even though this is not a
pixel-to-pixel correlation, it is clear that the enhanced coronal
emission occurs in a region with enhanced chromospheric activity.
The emission in the 304\,\AA\ band dominated by the \ion{He}{2} line
shows plasma below 10$^5$\,K and is still closely related to the
chromospheric network. The emission in the 131\,\AA, 171\,\AA, and 193\,\AA\ channels, which is dominated by plasma at 0.5\,MK
(\ion{Fe}{8}), 0.8\,MK (\ion{Fe}{9}), and 1.5\,MK (\ion{Fe}{12}), is
very much concentrated above one edge of the plage region. The image in the 211\,\AA\ channel (2.0\,MK; \ion{Fe}{14}) is almost identical to the 193\,\AA\ channel, which is why we do not include it here.

This enhanced coronal emission in the plage area could either be due
to small coronal loops within that region or it could originate from
footpoints of long hot loops rooted in that patch of enhanced
magnetic field. However, the 335\,\AA\  and
especially the 94\,\AA\ channels that should show hotter plasma -- if it would be
present -- do not show a signature of a long hot loop rising from
the patch in the center of the region. The 94\,\AA\
channel has two main contributions, one around
1.2\,MK (\ion{Fe}{10}) and another one at 7.5\,MK (\ion{Fe}{18}).
However, the emission pattern in the 94\,\AA\ channel is very similar to the 171\,\AA\ channel, which indicates that there is relatively little plasma reaching temperatures of $\approx$7.5\,MK along the line of sight. Likewise the 335\,\AA\ channel has main contributions around 0.1\,MK, 1\,MK, and 3\,Mm, but lacks a signature of a long hot loop, but shows more similarities to the 171\,AA channel (albeit much more noisy). 
For a detailed discussion of the temperature response of the AIA
channels see  \citep{Boerner+al:2012}.\footnote{One could also
speculate that the AIA short wavelength channels might show mainly
cool transition region emission; in particular the 193\,\AA\ and
211\,\AA\ channels have a significant contribution from temperatures
around 0.2\,MK (\ion{O}{5}) in quiet Sun conditions. However, then
these two channels should share some characteristics with the
304\,\AA\ channel, which they do not. In fact, the brightening in
193\,\AA\  and 211\,\AA\ is not overlapping at all with the pattern
in 304\,\AA. Thus we can conclude that the 193\,\AA\ and 211\,\AA\
channels really show plasma primarily at 1.5\,MK to 2\,MK.}

Unfortunately there is no X-ray image taken at the same time and location. However, an inspection of an image from the Hinode X-ray telescope \citep[XRT,][]{Golub+al:2007} taken about an hour before does not clearly resolve the issue. Therefore we assume that in the field of view of \fig{F:context.plage} the 193\,\AA\ emission in the center square is not a moss-type emission from a large hot loop.

Based also on Hi-C data, \cite{Testa+al:2013} see moss-type emission at the base of a hot ($>$5\,MK) loop, and find that it shows a high temporal variability. However, they look at a different part of the Hi-C field-of-view which is near a footpoint of a hot loop. In contrast, the brightening we discuss here is not related to a hot loop, as outlined above. 

If we can rule out that in our case the emission we see above the plage patch is
coming from the footpoint of long hot loops, it has to come from
small compact hot loops that close within the network patch. Such
small loops are clearly not resolved by the AIA images. However, in
the zoom to the Hi-C image of the plage area in \fig{F:plage}a we
can identify at least one structure that could be interpreted as a
miniature hot loop that reaches temperatures of (at least) about 1.5\,MK because we distinctively see it in the 193\,\AA\ channel of Hi-C (see arrows in panels a and c). This loop would
have a length (footpoint distance) of about only 1.5\arcsec\
corresponding to 1 Mm. For the width only an upper limit of 0.2 to
0.3\arcsec\ (150\,km to 200\,km) can be given, because the cross
section of this structure is barely resolved.

The HMI magnetogram in \fig{F:context.plage} shows only one polarity in the plage area, which would argue against a miniature coronal loop (following the magnetic field lines). However, it could well be that small-scale patches of the opposite polarity are found in this plage area, too. These small-scale opposite polarities would then cancel out in the HMI observations with its limited spatial resolution. At least high-resolution observations \citep[e.g.][]{Wiegelmann+al:2010} as well as numerical simulations of photospheric convection \citep[e.g.][]{Voegler+al:2005} show small-scale opposite polarities in otherwise largely monopolar regions with footpoint distances of the order of 1\,Mm or below. These small bipoles are the root of small flux tubes that have been found by \cite{Ishikawa+al:2010} by inverting high-resolution spectro-polarimetry data. They found this flux tube with a footpoint distance of about 1 Mm  to rise through the photosphere and then higher up into the atmosphere.
These miniature loops could be related to the short transition region loops that are found to reside within the chromospheric network \citep{Peter:2001:sec}.

From this discussion we can suggest that the small elongated structures we see in Hi-C above network and plage regions are in fact tiny small loops reaching temperatures of 1.5\,MK or more. These miniature coronal loops would have lengths (of their coronal part) of only around 1\,Mm. If this is indeed the case, they would be interesting objects: they would be so short, that they would barely stick out of the chromosphere!
Measuring from the photosphere, these loops would be only 5\,Mm long, with a 2\,Mm photosphere and chromosphere at each end.
A 5\,Mm long semi-circular loop has a footpoint distance of about 3\,Mm. Thus such miniature loops would span across just a single granule, connecting the small-scale magnetic concentrations in the intergranular lanes.

These miniature loops might be a smaller version of X-ray bright points, which are enhancements of the X-ray emission related to small bipolar structures \citep{Kotoku+al:2007}. Because of limitations of the spatial resolution of the X-ray observations (above 1\arcsec) no X-ray bright points have been observed that are as small as the miniature loop reported here. Short loops have been studied theoretically \citep[e.g.][]{Mueller+al:2003,Sasso+al:2012} indicating that such short hot structures are sensible, even though these short model loops show peak temperatures of well below 1\,MK.
\cite{Klimchuk+al:1987} found that hot short loops with heights below 1000\,km are thermally unstable and evolve into cool loops with temperatures around 10$^5$\,K. This would apply to the miniature loops proposed here, which would not be stable, anyway, because they can be expected to be disturbed rapidly by the convective motions of the granulation.

Future observational and modeling studies will have to show if this interpretation of miniature loops is correct, or the small-scale brightening in the plage region is better understood by the emission from the footpoint of a hot ({\small$\begin{array}{@{}c@{}}>\\[-1.5ex]{\sim}\end{array}$}3\,MK) loop. In particular the upcoming Interface Region Imaging Spectrograph \citep[IRIS;][]{Wuelser+al:2012}\footnote{See also http://iris.lmsal.com.} with its high spatial resolution (0.3\arcsec) covering emission originating from chromospheric to flare temperature to be launched in summer 2013 will be well suited for investigating this from the observational side.

\subsection{Substructure in large coronal loops\label{S:loops}}

We now turn to the discussion of the substructure of larger coronal loops. In \fig{F:roi} a loop system is visible that connects the periphery of the active region to the surrounding network. In the following we will concentrate on the small box labeled ``loops'' in \fig{F:roi}, but our results apply also to the other large loops in this figure. These loops show up also in AIA 171\,\AA\ images but are faint in the AIA 211\,\AA\ band, which hints at a temperature of the loops in the range of 1\,MK to 1.5\,MK.

We choose this particular area of the Hi-C field-of-view because it is the area  containing the most structures that can be easily identified as long coronal loops, having arc lengths of longer than about 50\arcsec\ (corresponding to 36\,Mm). The Northern part of the Hi-C field-of-view is dominated by plage and moss-type emission around the sunspots, as is clear from \figs{F:full} and \ref{F:chromo}. This upper part is thus dominated by more compact structures. In the Southern part of the image the region we selected shows the clearest loops. As mentioned also by \cite{Testa+al:2013}, the brightest emission seen by Hi-C is originating from the plage and moss areas. The longer loops presumably reaching higher altitudes might have a lower density and thus a smaller emission when compared to the moss that originates from the footpoints of hotter loops.

\begin{figure}
\includegraphics{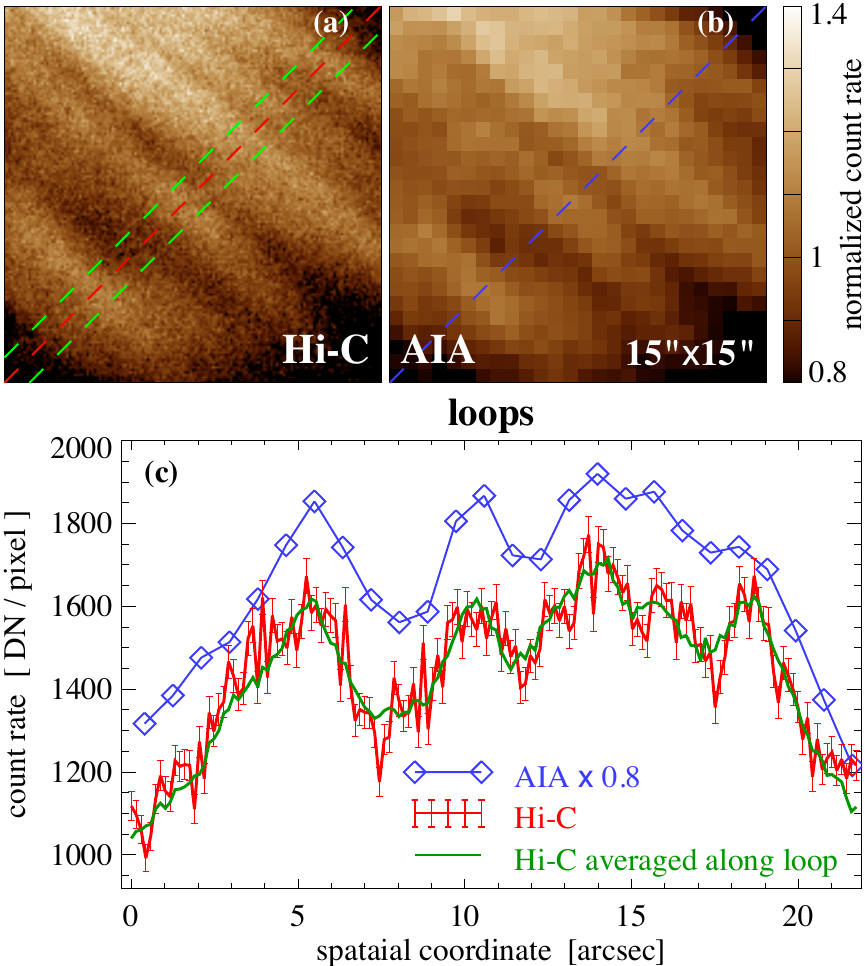}
\caption{Zoom of the \emph{loop region} indicated in \fig{F:roi} by a square. Otherwise this is the same as \fig{F:plage}. In addition, here the green line shows the cross-sectional cut averaged over 3\arcsec\ along the loops, i.e., it shows the average variation in the band defined by the two green lines in panel (a).
See \sect{S:loops}.
\label{F:loop}}
\end{figure}

In panels (a) and (b) of \fig{F:loop} we show a zoom into this loop region, where the loops are seen passing roughly along the diagonal. In the AIA image (panel b) the individual pixels are clearly identifiable; this field of view consists of only 26${\times}$26 pixels. In contrast, the Hi-C image (150${\times}$150 pixels; panel a) shows a larger degree of noise. This is due to the lower count rate per pixel because of the higher spatial resolution, i.e.\ the smaller pixel size, and the higher noise level of the Hi-C camera as compared to AIA. Still, from a comparison of the two images in panels (a) and (b) it is clear, that the Hi-C image does not show a coherent substructure of the loops that is aligned with the loops.

This missing substructure of the loops in \fig{F:loop} becomes evident when investigating the cut perpendicular to the loops shown in panel (c) of \fig{F:loop}. If one would subtract the background, the loops in AIA would have a width of some 4 pixels, i.e.\ they would be very close to the resolution limit of AIA. This size corresponds to some 1.8\arcsec\ to 2.4\arcsec\ or 1.3\,Mm to 1.7\,Mm. The cut of the Hi-C image confirms this width, which is very clear for the loops at spatial positions 5\arcsec\ and 10\arcsec. Looking at AIA alone, one might have missed the small structure at 18\arcsec\ which is at the side of a bigger one. These results are consistent quantitatively and qualitatively with \cite{Brooks+al:2013} who looked at short segments of a larger number of (long and short) loops.

Most importantly, the Hi-C data in panel (c) of \fig{F:loop} do not
show an indication of a substructure of the loops that is sticking
out of the noise (the Poisson errors are plotted as bars). When
plotting the cross section of the loops not as a single-pixel cut,
but averaged along the loops, the noise disappears and the loop
cross-sections are smooth. The green line shows the variation in the
3\arcsec\ wide band defined by the green dashed lines in panel (a).
This averaging over 21 Hi-C pixels reduces the (Poisson) noise by a
factor of about 4.5, giving a reduced error of about 20 counts.\footnote{Alternatively one could have averaged in time or in both time and space. For the case at hand this averaging along the loops seems appropriate because the loops in \fig{F:loop} show a nice smooth variation along the loop.} This
just corresponds to the variability still seen in the averaged
cross-sectional cut. In Appendix \ref{S:fft} we give some further details on the size of the loops seen in Hi-C.

From the above we can conclude, that within the instrumental capabilities of Hi-C no plausible substructure of the long coronal loops can be seen, but the long loops are smooth and resolved. The fact that we see structures in the Hi-C image in the plage area (\sect{S:plage}) that are much smaller than the loop cross-sections reassures us that the spatial resolution of Hi-C is sufficient to see a substructure, if it both existed and were bright; and that the averaging process would reveal the structure if it were parallel to the arcsec-scale strands in the loops.

Of course, this observation does not rule out the possibility that the loops might have a substructure on scales smaller than observable by Hi-C. Still, with these Hi-C observations we can set an upper limit for the diameter of individual \emph{strands} that might compose single loops.

\subsection{Upper limit for the diameter of strands in loops\label{S:strands}}

In the following we will estimate an upper limit for the thickness of individual strands composing a coronal loop. For this we will use the argument that the emission seen across the loop, i.e. the cross-sectional cut, should be smooth, just as found in the observations. We will assume that all strands are circular in cross section and run in parallel. This is the simplest model possible, and certainly reality will be much more complicated. However, for a rough estimation these assumptions should suffice.

\begin{figure}
\includegraphics{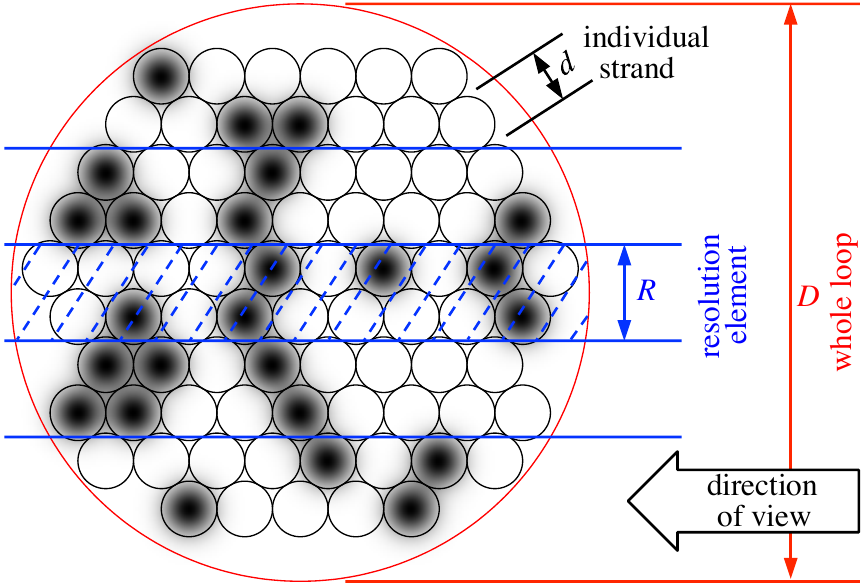}
\caption{Cartoon of the multi-stranded loop. The loop with diameter
$D$ is composed of many individual strands with diameter $d$. When
observing the loop, each spatial resolution element of size $R$ of
the instrument corresponds to a column of the cross section. The
hashed area represents one such column. Strands that are bright in a
particular EUV channel (or line) are indicated by a back dot, empty
circles represent strands not radiating in this channel (at this
instant in time). This example consists of $N_t{=}79$ strands in
total, of which $N_b{=}25 $ are bright, corresponding to a fraction
$f_b{=}0.3$. For the loops on the Sun we estimate that they consist
of $N_t{\approx}2500$ strands.
See \sect{S:strands}.
\label{F:strands}}
\end{figure}

We start with a loop with diameter $D$ that is composed of a total number $N_t$ of strands, each of which has the same diameter $d$ (see \fig{F:strands}).
Of all the strands only a fraction $f_b$  is bright, so that the number of bright strands is
\begin{equation}\label{E:num.bright}
N_b = N_{t}~f_b ~.
\end{equation}
This fraction $f_b$ is equivalent to the fraction of time each
individual strand with time-dependent heating and temperature and
density structure will be visible in a specific EUV line or channel.
Based on multi-stranded loop models this fraction can be estimated
to have an upper limit of $f_b\approx0.1$ \citep[e.g.][]{Warren+al:2003,Warren+al:2008,Viall+Klimchuk:2011}.
The fraction $f_b$ is also equivalent to the volume filling factor of the (bright) plasma in the corona. In observations of bright points \cite{Dere:2009} found values in the range of 0.003 to 0.3 with a median value of 0.04, which support our choice for an upper limit. One should remember, that the filling factor might be much smaller, in particular when considering cooler plasma. For the transition region \cite{Dere+al:1987} found filling factors ranging from 0.01 down to $10^{-5}$.

When observing with an instrument with a spatial resolution $R$, each resolution element will represent a column of a cut through the loop (cf. \fig{F:strands}). In this \emph{column} there are $N_c$ bright loops. In order to have a smooth cross-sectional profile we require that neighboring resolution elements contain a similar number of bright strands (each of the same brightness). The difference of the number of strands in neighboring resolution elements should then follow Poisson statistics, $\mathit{\Delta}N_c\approx\sqrt{N_c}$. The relative difference of the number of strands in neighboring resolution elements directly gives the brightness variation, and for a smooth profile we require this to be smaller than $\varepsilon$,
\begin{equation}\label{E:num.column}
\frac{\mathit{\Delta} N_c}{N_c} < \varepsilon
\quad\to\quad
N_c > \frac{1}{\varepsilon^2} ~.
\end{equation}
Typically one would require $\varepsilon{\approx}0.1$, i.e., a pixel-to-pixel variation across the loop of less than 10\% for a smooth profile.

From the number of bright strands in one column, $N_c$, we can estimate the number of bright strands in the whole loop, $N_b$, by the ratio of the cross section of the whole loop and of a single column {}(e.g. hashed column in \fig{F:strands}), which together with \eqn{E:num.column}
yields
\begin{equation}\label{E:num.bright.ii}
N_{b} = \frac{\pi\,(D/2)^2}{R\,D}~N_c
\quad\to\quad
N_b > \frac{\pi}{4}~\frac{D}{R}~\frac{1}{\varepsilon^2} ~.
\end{equation}
Assuming that the strands in the loop are packed as dense as possible, the cross section of the loop as a whole and of the $N_t$ strands are related by
\begin{equation}\label{E:densest.packing}
\pi\left(\frac{D}{2}\right)^{\!\!2} = \frac{\pi}{\sqrt{12}} ~ N_t ~ \pi\left(\frac{d}{2}\right)^{\!\!2} ~,
\end{equation}
where the factor of $\pi/\sqrt{12}\approx0.9$ stems from the densest packing of circles in a plane.
Using \eqn{E:num.bright} and (\ref{E:num.bright.ii}) this now gives the upper limit for the diameter of individual strands,
\begin{equation}\label{E:strand.limit}
d ~~ < ~~ \frac{2\sqrt{2}\sqrt[4~]{3}}{\pi} ~~ \varepsilon ~ \sqrt{f_b} ~~ \sqrt{R\,D} ~~~ \approx ~~~ 1.2 ~ \sqrt{f_b} ~ \varepsilon ~\sqrt{R\,D}   ~,
\end{equation}
and the lower limit for the total number os strands in the loop,
\begin{equation}\label{E:strand.N}
N_t ~~ > ~~ \frac{\pi}{~4~} ~ \frac{D}{~R~} ~ \frac{1}{~f_b~\varepsilon^2~} ~.
\end{equation}

For the loops observed here with a diameter of $D{\approx}2$\arcsec\ and the resolution of Hi-C of $R{\approx}0.2$\arcsec\  assuming a fraction of the bright loops of $f_b{\approx}0.1$ and a pixel-to-pixel variation of $\varepsilon{\approx}0.1$ in the observation, we find an upper limit for the diameter of an individual strand of $d{\approx}0.025$\arcsec\ corresponding to about
\begin{equation}\label{E:strand.upper.limit}
d < 15\,{\rm{km}}~.
\end{equation}
This strand diameter of only 15\,km is small compared to the loop diameter --- this loop would have to host at least about $N_t{\approx}7500$ strands of which $N_b{\approx}750$ are bright in the given EUV channel at any given time. In Appendix \ref{S:num} we discuss a simple numerical experiment confirming this conclusion.  If one would adopt the lower value of 0.003 for the filling factor derived by \cite{Dere:2009}, one would end up with a quarter million strands with diameters of only 3\,km.

Strictly speaking, this discussion applies only for the 193\,\AA\ channel. Other bandpasses respond differently to different heating scenarios in loop models \citep[e.g., see the review of][]{Reale:2010} and thus might show different filling factors. Still, our results can be expected to apply roughly for emission originating from the coronal plasma at 1\,MK to 2\,MK.

Together with the discussion in \sect{S:loops} we conclude that the loops are either monolithic structures, the diameter of the individual strands has to be smaller than 15 km, or else the strands must be implausibly well organized. This new upper limit is more than a factor of 10 smaller than derived from previous studies \citep[e.g.][]{Brooks+al:2012}, which became possible by the enhanced spatial resolution of Hi-C. The multi-stranded loop scenario can only be valid if the upper limit of the strands set by observations is larger than the lower limit for the strand diameter set by basic physical processes such as reconnection, gyration, heat conduction or turbulence. At this point we think that this is the case, however more detailed studies, in particular of MHD turbulence, would be needed for a final conclusion. We discuss these issues briefly in \app{S:lower.limit}.

\section{Loop morphology and comparison to a 3D model\label{S:model}}

In a first attempt for a morphological comparison between the Hi-C observations and a 3D coronal model we will just highlight some common features between observation and model on a qualitative basis. For final conclusions, a more detailed quantitative comparison is needed, and in particular a more in-depth analysis of the model is required to better understand how the various structures do form.

\begin{figure}
\includegraphics{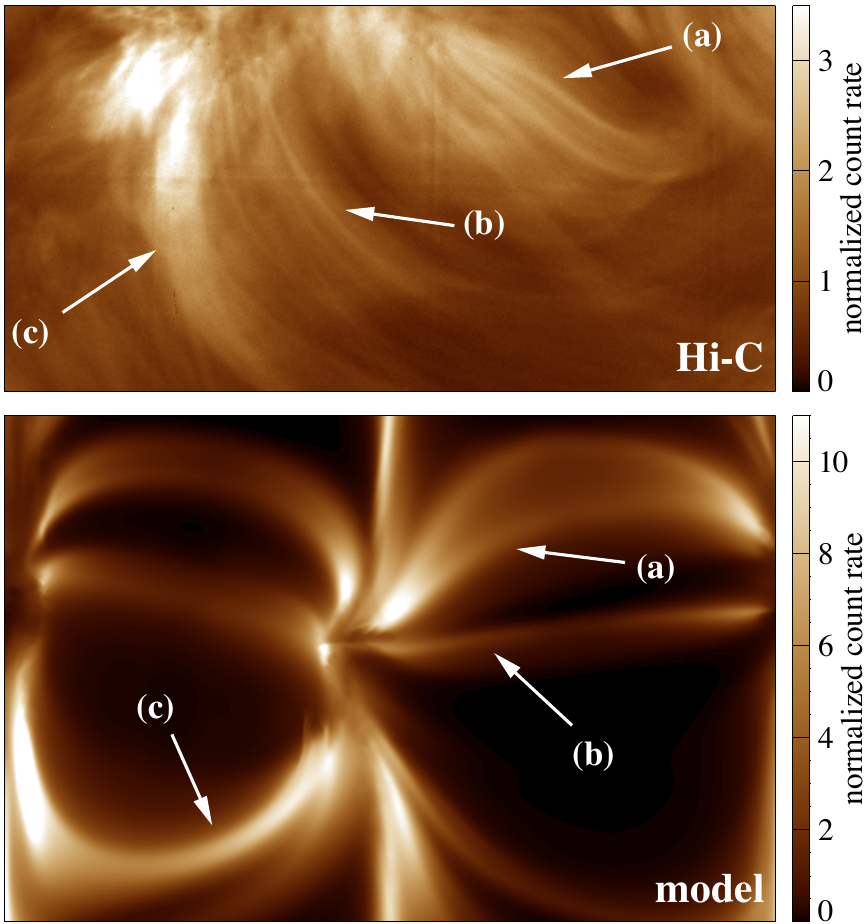}
\caption{Morphological comparison of observation and model.
The top panel shows the actual observation of Hi-C (193\,\AA\ band). The field of view (124\arcsec$\times$62\arcsec) 
is outlined in \fig{F:full} by the dashed rectangle. The bottom panel shows the coronal emission as synthesized from a 3D MHD model for this channel ($165{\times}109$\,Mm). The arrows point to features that can be found in both model and observations: (a) expanding envelope that consists of several non-expanding loops, (b) thin non-expanding threads, and (c) rather broad loop-like structures with approximately constant cross section.
See \sect{S:model}.
\label{F:model}}
\end{figure}

The coronal loops in the field-of-view of the Hi-C observations discussed in this manuscript can be classified (by eye) into three categories (cf. arrows in \fig{F:model}, top panel):
\begin{itemize}
\item[(a)] Expanding envelope that
consists of several non-expanding loops.
\item[(b)] Thin ({\small$\begin{array}{@{}c@{}}<\\[-1.5ex]{\sim}\end{array}$}3\arcsec) non-expanding threads.
\item[(c)] Broad ({\small$\begin{array}{@{}c@{}}>\\[-1.5ex]{\sim}\end{array}$}5\arcsec) loop-like structures with approximately constant cross
section.
\end{itemize}
The individual loops in the expanding structure (a) and the thin
threads (b) seem to show no (significant) expansion. We are
investigating this quantitatively with the Hi-C data and will
present our results in a separate paper.  The tendency for
constant cross-section was reported more than a decade ago for loops observed
in the EUV and X-rays \citep{Klimchuk:2000,Watko+Klimchuk:2000,LopezFuentes+al:2006}.
Recently, based on a 3D MHD model, it has been shown that the constant cross
section could be due to the temperature and density structure within
in the expanding magnetic structure, in interplay with the formation
of the coronal emission lines \citep{Peter+Bingert:2012}. This would
work for EUV observations, but it still has to be investigated if
this would work also at X-ray wavelengths which originate from a
much broader range of high temperatures.

Recently, \cite{Malanushenko+Schrijver:2013} have suggested that the constant cross section result may be
an artifact of the observing geometry and the likelihood that the
shape of the cross section varies along the loop.  The cross
sectional area could expand, but if it does so preferentially along
the line-of-sight, then the loop thickness in the plane-of-the sky
(i.e., the image) will be constant.  This could certainly explain
many loops.  However, there should be many other cases where a
different observing geometry reveals a very strong expansion.  Such
cases need to be verified before we can accept this explanation for
the constant cross section loops. Also, the thermal structure along and perpendicular to the loops has to be considered, and not the magnetic structure alone, as pointed out by \cite{Mok+al:2008} and \cite{Peter+Bingert:2012}.

Because the 3D MHD\ model successfully provided a match to the constant cross-section loops, we compared a snapshot of a 3D MHD model to the Hi-C observation to see if we find the three categories of loop structures also in the emission synthesized from the model. In the 3D MHD we solve the mass, momentum and energy balance in a box spanning  167${\times}$167\,Mm$^2$ horizontally and 80\,Mm in the vertical direction (512$^3$ equidistant grid). Horizontal motions (of the granulation) drive the magnetic field in the photosphere and lead to braiding of magnetic field lines as originally suggested by \cite{Parker:1972}. This process induces currents that are dissipated in the upper atmosphere and by this heat the corona. The details of this model have been described by \cite{Bingert+Peter:2011,Bingert+Peter:2013}.
The numerical experiment we use here differs from the \cite{Bingert+Peter:2011,Bingert+Peter:2013} model by the magnetic field at the lower boundary, an increased size of the computational domain and a higher spatial resolution. As described by \cite{Peter+al:2006} we interpolate the MHD quantities to avoid aliasing effects and compute the emission as it would be seen by AIA using the temperature response functions \citep{Boerner+al:2012,Peter+Bingert:2012}. In \fig{F:model} we show the synthesized emission in the 193\,\AA\ channel\footnote{Because of a slightly too high density in the model transition region we reduce the density (by up to a factor of 2) there in order to avoid small-scale artifacts due to the contribution of plasma at low temperatures.} integrated along the vertical axis, i.e. when looking from straight above.

In the image synthesized from the model we find the same three classes of structures as in the actual observation; the labeled arrows point at such structures. While the thin constant-cross-section loops (b) have been discussed before \citep{Peter+Bingert:2012}, here we also find the non-expanding loops in an expanding envelope (a) and the broad loop-like structures (c). It is beyond the scope of this (mainly observational) study to perform a detailed analysis of the 3D MHD model to investigate how exactly what the nature of these three categories is. Here we simply state that these categories are also found in numerical experiments, so that there is some hope to  understand how they come about in future studies.

\section{Conclusions\label{S:conclusions}}

In this study we presented results on the structure of coronal loops based on new observations with the Hi-C rocket telescope providing unprecedented spatial resolution in the EUV down to 0.2\arcsec\ at a spatial scale of 0.1\arcsec\ per pixel.

We have found miniature loops hosting plasma at 1.5\,MK with a length of only about 1\,Mm and and a thickness of below 200\,km (\sect{S:plage}). With other current instrumentation, such as AIA, these would cover just two spatial pixels.
These miniature loops are consistent with small magnetic flux tubes that have been observed to rise through the photosphere into the upper atmosphere. However, it will be a challenge to understand these miniature loops  in terms of a (traditional) one-dimensional model. From the observational side, this clearly shows the need for future high-resolution Hi-C-type observations together with high-resolution spectro-polarimetric observations of the photosphere and chromosphere.

In the case of the longer more typical coronal loops we found that the Hi-C observations do not show indications of a sub-structure in these loops (\sect{S:loops}). Therefore these loops with diameters of typically 2\arcsec\ to 3\arcsec\ are either real monolithic entities or they would have to  be composed by many strands with diameters well below the resolution limit of Hi-C. Based on some simple assumptions we found that the strands would have to have a diameter of at most 15\,km, which would imply that a loop with 2\arcsec\ diameter would have to be composed of at least 7500 individual strands (\sect{S:strands}). This would compare to a 1\,cm diameter wire rope consisting of wire strands of only 0.1\,mm diameter. No matter if the loops are monolithic or multi-stranded in nature, it still remains puzzling what determines the width of the loop of typically 2\arcsec\ to 3\arcsec, which is found consistently in the Hi-C as well as the AIA data.
The observational time and field-of-view of the Hi-C rocket experiment were limited, so this discussion cannot be generalized. This highlights the need for such high-resolution observations of the corona in future space missions.

Numerical experiments show similar (large) loop structures as found in the Hi-C observations: non-expanding loops in expanding envelopes, thin threads and thick constant cross-section loops. It still needs to be determined how these are produced in the numerical experiments, and if the processes in the model can be realistically applied to the real Sun. Anyway, the 3D numerical experiments provide us with a tool to investigate this in future studies and thus learn more about the nature of loops in the corona, miniature and large.

{
\acknowledgements 
We acknowledge the High resolution Coronal Imager instrument grant funded by the
NASA's Low Cost Access to Space program. MSFC/NASA led the mission and partners
include the Smithsonian Astrophysical Observatory in Cambridge, Mass.; Lockheed Martin's
Solar Astrophysical Laboratory in Palo Alto, Calif.; the University of Central Lancashire in Lancashire, England; and the Lebedev Physical Institute of the Russian Academy of Sciences in Moscow.
The AIA and HMI data
used are provided courtesy of NASA/SDO and the AIA and HMI science
teams. The AIA and HMI data have been retrieved using the German
Data Center for SDO. The numerical simulation was conducted at the
High Performance Computing Center Stuttgart (HLRS). This work was partially funded by the Max-Planck/Princeton Center for Plasma Physics.
The work of J.A.K. was supported by the NASA
Supporting Research and Technology and Guest Investigator Programs.
H.P. acknowledges stimulating discussions with Robert Cameron and Aaron Birch.
We thank the anonymous referee for constructive comments.
}




\vspace{2cm}

\appendix

\section{Lower limit for the strand diameter\label{S:lower.limit}}

The typical length scale \emph{perpendicular} to the magnetic field through whatever process places a \emph{lower} limit for the strand diameter $d$. If $d$ would be smaller than this length scale, the neighboring strand would interact through this process and consequently the strands would no longer be individual structures but one common entity.

This discussion is of interest, because if this lower limit for the strand diameter would be larger than the upper limit derived from observations in \sect{S:strands}, the loops would have to be monolithic structures.

\paragraph{Reconnection.}
Most probably reconnection is at the basis of the heating process. For reconnection to happen the inductive term and the dissipative term in the induction equation have to be of the same order, or in other words, the magnetic Reynolds number has to be of order unity, i.e., $Rm=U\,\ell/\eta\approx 1$. Here $U$ is the typical velocity and $\eta$ is the magnetic diffusivity. The length scale $\ell$ would represent the thickness of the resulting current sheet. This $\ell$ is then a lower limit for the strand diameter. If strands would be thinner, they would be part of the same reconnection region and thus not distinguishable.

Following \cite{Spitzer:1962} one cane derive the electric conductivity $\sigma$ from classical transport theory, which then provides $\eta=(\mu_0\,\sigma)^{-1}$. In the corona at $10^6$\,K the value is $\eta\approx1\,{\rm{m}}^2\,{\rm{s}}^{-1}$. For $Rm{\approx}1$ the length scale is given by $\ell\approx\eta/U$. For a small value of $U{\approx}1$\,km/s (certainly there are much faster flows in the corona), we find a value of
\begin{equation}\label{E:length.rec}
\ell \approx 1\,{\rm{mm}} ~.
\end{equation}
Arguments along this line of thought can be found in e.g. \cite{Boyd+Sanderson:2003} or in the preface of \cite{Ulmschneider+al:1991}. This value must be a vast underestimation, because it is much smaller than the ion (and even the electron) gyro radius.

\paragraph{Gyration.}
Another relevant length scale is the Larmor or gyro radius of the gyration of the ions and electrons in the corona. This is given by $r_{\rm{L}} = m\,v_\perp (e\,B)^{-1}$ , with the mass $m$ and the charge $e$ of the particles and the magnetic field $B$. Assuming that the perpendicular velocity is given by the thermal speed, $v_\perp=(3kT/m)^{1/2}$, one finds
\begin{equation}\label{E:length.gyro}
r_{\rm{L}} = \frac{~(3kTm)^{1/2}~}{eB}
\qquad\to\qquad
r_{\rm{L,p}} \approx 1.5\,{\rm{m}}  ~.
\end{equation}
Thus in the corona at temperatures of about $T{\approx}10^6$\,K and for $B{\approx}10$\,G the Larmor radius of the protons is of the order of 1\,m.

\paragraph{Heat conduction.}
The head conduction \emph{perpendicular} to the magnetic field is much less efficient than the parallel conduction, but if the temperature gradients across the field become large enough it might become non-negligible. For the coefficient of the perpendicular heat conduction one can derive \citep[e.g.][Sect.\,2.3.2]{Priest:1982}
\begin{eqnarray}
\nonumber
\kappa_\perp =
          3.6\,10^{-8}\, \frac{\rm W}{\rm ~K~m~}  ~
    \frac{~ \big(n\,[10^{15}\,{\rm{m}}^{-3}]\big)^2 ~~
          }{~\big(T\,[10^{6}\,{\rm{K}}]\big)^{1/2}
            ~~ \big(B\,[10\,{\rm{G}}]\big)^{2}
            ~~ \ln\mathit{\Lambda} ~} ~,
\\ \label{E:heat.perp}
\end{eqnarray}
where $n$ is the particle density and in the corona the Coulomb logarithm is roughly $\ln\mathit{\Lambda}{\approx}13$.

We now consider the minimum diameter $d$ of a strand with length $L$. For this we assume that the energy lost by heat conduction perpendicular to the magnetic field, $\kappa_\perp\,{\nabla}T$, through the mantle surface of the strand, $L\,{\pi}d$, is balanced by the energy input $F_H$ through the two footpoints of the strand with cross section ${\pi}d^2\!/4$,
\begin{equation}\label{E:heat.eq}
\kappa_\perp~\nabla T ~ L\,\pi d ~~=~~ 2~F_H ~ \pi \frac{d^2}{4}
\end{equation}
Using expression (\ref{E:heat.perp}) one finds
\begin{eqnarray}
\nonumber
d ~>~ 23\,{\rm{m}}~~\frac{~ n\,[10^{15}\,{\rm{m}}^{-3}] ~~
            \big(T\,[10^{6}\,{\rm{K}}]\big)^{\!1/4} ~~
            \big(L\,[100\,{\rm{Mm}}]\big)^{\!1/2}
          }{B\,[10\,{\rm{G}}] ~~ \big(F_H\,[1000\,{\rm{W}}\,{\rm{m}}^{-2}]\big)^{\!1/2}}
          ~.
\\
\label{E:lower.limit}
\end{eqnarray}
For typical coronal values and an energy flux density into the corona of $1000\,{\rm{W}}\,{\rm{m}}^{-2}$, we find a diameter of the order of 20\,m. This is a lower limit for the strand diameter. If the strand would be thinner, the temperature gradient to the neighboring (cold) strand would be higher, and thus the energy loss by perpendicular heat conduction would be stronger than the energy input. Consequently the strand would start to dissolve into the neighboring strands increasing its diameter.

\paragraph{Other processes.}
From the above discussion we find that the strands should be at least some 10\,m to 50\,m in diameter, set by the perpendicular heat conduction. However, this lower limit might be on the small side. Other processes, in particular MHD turbulence, might increase the length scale perpendicular to the magnetic field considerably. Then the reconnection process will effectively operate on larger length scales, and also the heat conduction perpendicular to the magnetic field will be more effective, ensuring smaller temperature gradients and thus larger length scales perpendicular to the magnetic field. The MHD turbulence simulations of \cite{Rappazzo+al:2008} show elongated current concentrations along the magnetic field, which could be interpreted as the strands in a loop. However, because of the lack of heat conduction and radiative losses in their model, which would have been beyond the scope of their study, one cannot say much on the diameter of the potentially developing strands.

Future theoretical investigations might place a better lower limit to the strand diameter and and thus further limit the range of possible strand diameters given by the relations (\ref{E:strand.upper.limit}) and (\ref{E:lower.limit}). Finally such studies would help in deciding wether loops are monolithic or multi-stranded.

\section{Size of the loops seen in Hi-C\label{S:fft}}

\begin{figure}
\includegraphics{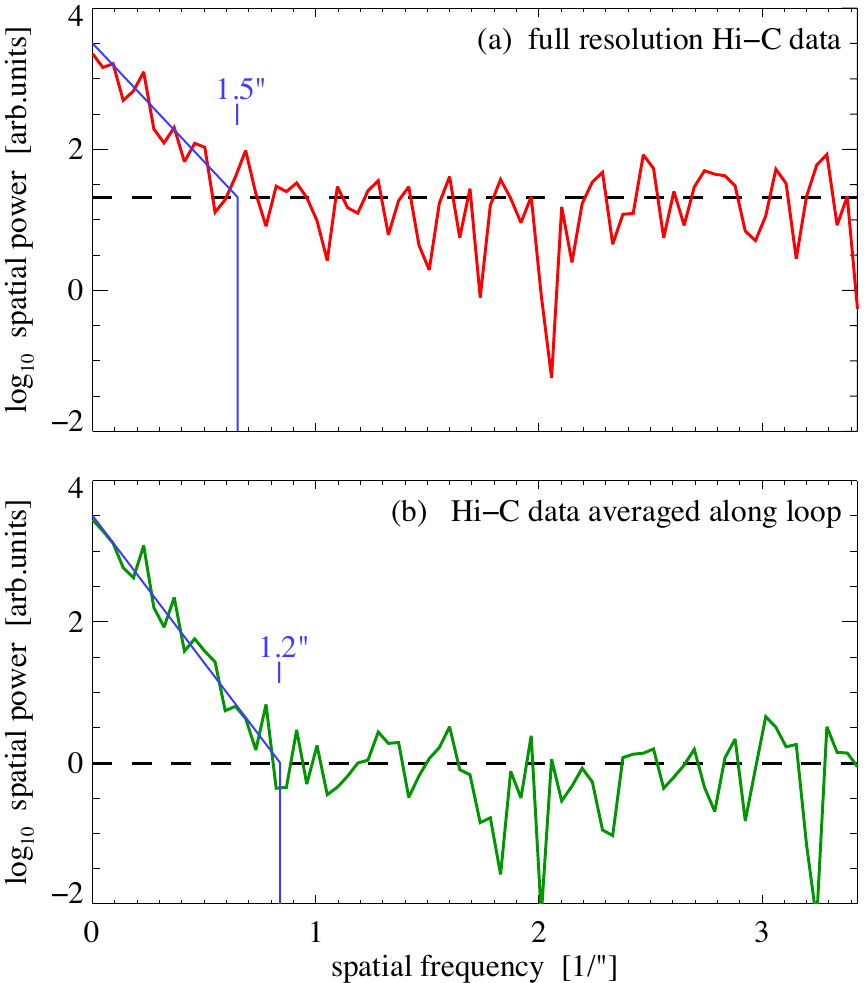}
\caption{Size of the loop structures in Hi-C.
Shown are the power spectra of the cross-sectional cuts of the loops in \fig{F:loop}, for the original data (top panel, red here and in \fig{F:loop}) and the data averaged over 21 pixels along the loop (bottom panel green here and in \fig{F:loop}). The horizontal dashed lines shows the noise level (21 in top panel, 1 in bottom panel). The diagonal blue line is a by-eye fit to the power spectrum at small spatial frequencies, the vertical line indicated the intersection of this fit with the respective noise level. The numbers in the panels give the corresponding spatial scale.
See Appendix \ref{S:fft}.
\label{F:fft}}
\end{figure}

To obtain a quantitative estimate for the size of the loop structures discussed in \sect{S:loops} we perform a Fourier transform of the spatial variation shown in \fig{F:loop}. We do this for the original profile across the loop (red with bars in \fig{F:loop}) as well as for the data averaged 21 pixels along the loops (in green).
After subtracting the linear trend and apodizing using a Welsh filter we perform a Fourier transform to obtain the power spectrum as a function of spatial frequency. The resulting power spectra are shown in \fig{F:fft}. The top panel shows the power spectrum for the original profile, the bottom panel for the averaged data.

We calculate the noise level in the power spectrum by equating the square of the error in the data integrated over space to the noise in the power integrated over spatial frequency. For the data averaged aver 21 pixels along the loop we assumed the error to be given by the original errors scaled down by $\sqrt{21}$. We overplot these noise levels for the original and the spatially averaged data in \fig{F:fft} as dashed lines.

From the power spectra in \fig{F:fft} it is clear that all structures at scales below 1.2\arcsec\ are consistent with noise (i.e., right to the vertical blue lines). The corresponding frequency of 0.8\,1/\arcsec\ is far from the Nyquist frequency of about 3.4\,1/\arcsec. The results are similar for the original and the averaged Hi-C data. Based on the averaged Hi-C data, which show less noise, the structures at scales below 1.2\arcsec\ represent less than 1\% of the power. If there would be significant power from (sub-)structures in the loops investigated in \fig{F:loop} it should show up somewhere in the frequency range from 1\,1/\arcsec\ to 3.4\,1/\arcsec, where Hi-C would be sensitive.

In other parts of the field-of-view for different structures, e.g. in moss areas or for the miniature loop discussed in \sect{S:plage}, Hi-C shows much smaller structures basically down to its resolution limit.
However, the loop structures investigated in \sect{S:loops} and \fig{F:loop}, which have widths of about 2\arcsec, do not show a substructure on scales below 1.2\arcsec. This is equivalent to saying that the long 2\arcsec\ wide coronal loops as seen in Hi-C have no substructure.

\section{Appearance of multi-stranded loops\label{S:num}}

\begin{figure*}
\centerline{\includegraphics{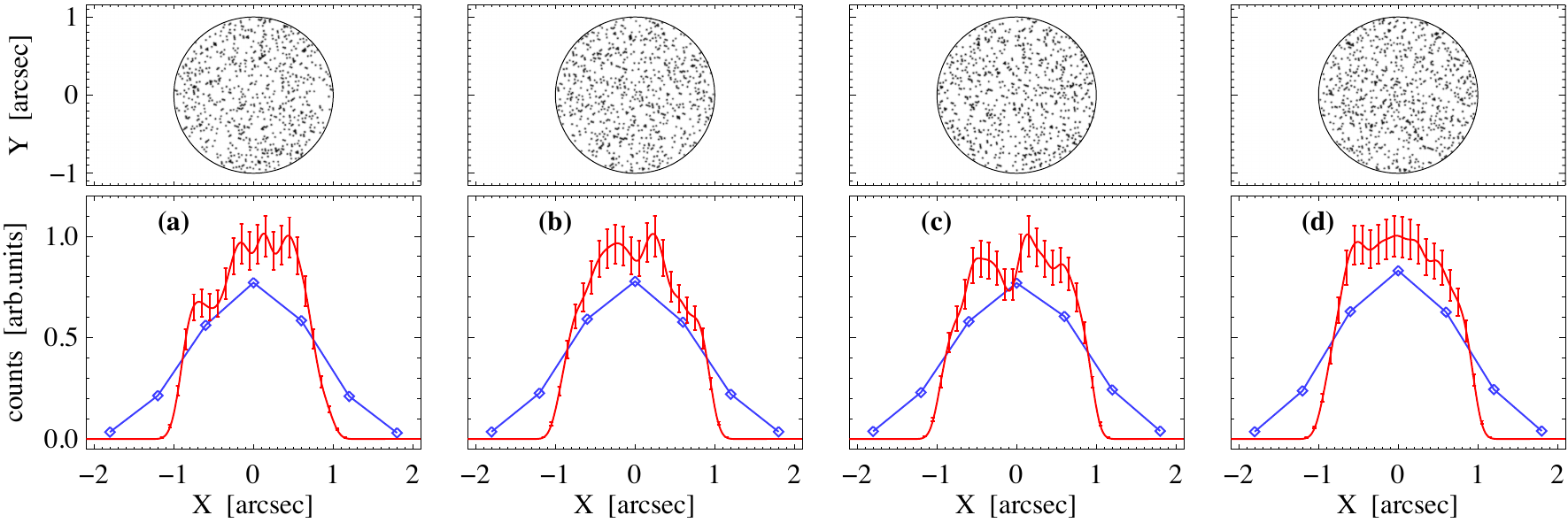}}
\caption{Numerical experiments for the cross-sectional profile of multi-stranded loops. The top panels show the loop cross-section for four experiments with about 7500 strands of which about 750 are bright. They differ only by the random selection which strands are bright. Each strand has a diameter of 15\,km. The bottom panels show the respective cross-sectional profiles (i.e., integration along Y) as it would be observed by AIA (blue, diamonds) and Hi-C (red). The height of the bars for Hi-C correspond to 10\% of the counts. See Appendix\,\ref{S:num}.
\label{F:num}}
\end{figure*}

To visualize the analytical results in \sect{S:strands} on the upper limit for the strand diameter and the lower limit for the number of bright strands in the loop we conduct a simple numerical experiment. In the following all variables have the same meaning as in \sect{S:strands}.

For the experiment we assume each individual strand to have a Gaussian cross-sectional profile (in intensity), where the full width at half maximum is used to define the strand diameter $d$. We then fill the circular loop cross section with a most dense packing of circles, which provides us with the total number of strands $N_t$. For the numerical experiment we select randomly which of the loops are bright, with a fraction $f_b$ of all loops being bright, $N_b=N_t\,f_b$.

We then integrate along a line-of-sight perpendicular to the loop, which provides the cross-sectional intensity profile of the loop. This profile we fold with a point-spread-function for AIA and Hi-C (where for simplicity we assumed a Gaussian profile which is sufficient for the purpose at hand).

In \fig{F:num} we visualize the results for the parameters as found in \sect{S:strands}, i.e., a total of $N_t{=}7500$ strands of which a fraction of $f_B{=}0.1$ or $N_b{=}750$ is bright with a diameter of $d{=}15$\,km. These fit into a circular loop with a diameter $D$ corresponding to 2\arcsec\ or 1500\,km.

In the top panels of \fig{F:num} we show the cross section of the loop for four different random selections of which loops are bright. The respective lower panel shows the cross-sectional profile if it would be observed with AIA or Hi-C. It is clear that the limited spatial resolution of AIA of slightly worse than 1\arcsec\ does not allow to see any features. We see a smooth profile which is basically identical for all four cases.

The Hi-C data show some variation in the cross-sectional profile. In \sect{S:strands} we used the variability $\varepsilon$ to derive the upper limit of the strand diameter. Here we plot the variability $\varepsilon$ as bars on top of the profile. We see that the variability in the simulated cross-sectional profile for the four cases in \fig{F:num} is roughly comparable with the variability $\varepsilon$ we used in \sect{S:strands}. If we would use a significantly larger (smaller) number of individual strands, we would find a smoother (rougher) cross sectional profile, which confirms the derivation of the limit for the strand diameter in \sect{S:strands}. Actually, the cross sectional profiles shown in \fig{F:num} look quite similar to the profiles of the loops seen in observations in \fig{F:loop}.

This numerical experiment also elucidates on the interpretation of substructures in loops. In the four examples we show in \fig{F:num} only one (d) shows a perfectly smooth profile. One might be tempted to conclude from a cross-sectional profile as in case (c) that this loop shows only two major substructures. In cases (a) and (b) one might underestimate the diameter of the loop. Looking at more randomly generated distributions of bright strands, one finds a large variety of cross-sectional profiles. This shows that one has to be careful in the analysis of single loops, because the (random) distribution of a large number of loops might mimic the existence of structures that are not present in reality. 

These numerical experiments confirm our conclusions based on the analytical analysis in \sect{S:strands} that the strand diameter should be of the order of about 15\,km or less (in particular when conducting experiments for loops with different numbers of strands, which we do not show here for brevity). Of course, this does not exclude the possibility that there is no substructure at all, but that the loop is a monolithic structure as noted in \sect{S:strands}.


\end{document}